\documentclass{article}

\usepackage{arxiv}

\usepackage[utf8]{inputenc} 
\usepackage[T1]{fontenc}    
\usepackage{hyperref}       
\usepackage{url}            
\usepackage{booktabs}       
\usepackage{amsfonts}       
\usepackage{nicefrac}       
\usepackage{microtype}      
\usepackage{lipsum}
\usepackage{graphicx}
\usepackage{siunitx}
\usepackage[super]{nth}
\usepackage{amsmath,amssymb,amsfonts, mathtools}
\usepackage{algorithmic}
\usepackage{stfloats}
\usepackage{textcomp}
\usepackage{xcolor}
\usepackage{soul}

\usepackage{enumitem}
\usepackage[ruled]{algorithm2e}
\usepackage[style=ieee,isbn=false,url=false,eprint=false,maxcitenames=1,mincitenames=1]{biblatex}

\title{Self-Supervised Learning via VICReg Enables Training of EMG Pattern Recognition Using Continuous Data with Unclear Labels}

\author{
 Shriram Tallam Puranam Raghu \\
  Department of Electrical and Computer Engineering and the Institute of Biomedical Engineering\\
    University of New Brunswick\\
    Fredericton, NB, Canada, E3B 5A3\\
   \And
 Dawn T. MacIsaac \\
  Department of Electrical and Computer Engineering and the Institute of Biomedical Engineering\\
    University of New Brunswick\\
    Fredericton, NB, Canada, E3B 5A3\\
  \And
 Erik J. Scheme \\
  Department of Electrical and Computer Engineering and the Institute of Biomedical Engineering\\
    University of New Brunswick\\
    Fredericton, NB, Canada, E3B 5A3\\
}

\begin{document}
\maketitle
\begin{abstract}
In this study, we investigate the application of self-supervised learning via pre-trained Long Short-Term Memory (LSTM) networks for training surface electromyography pattern recognition models (sEMG-PR) using dynamic data with transitions. While labeling such data poses challenges due to the absence of ground-truth labels during transitions between classes, self-supervised pre-training offers a way to circumvent this issue.  We compare the performance of LSTMs trained with either fully-supervised or self-supervised loss to a conventional non-temporal model (LDA) on two data types: segmented ramp data (lacking transition information) and continuous dynamic data inclusive of class transitions. Statistical analysis reveals that the temporal models outperform non-temporal models when trained with continuous dynamic data. Additionally, the proposed VICReg pre-trained temporal model with continuous dynamic data significantly outperformed all other models. Interestingly, when using only ramp data, the LSTM performed worse than the LDA, suggesting potential overfitting due to the absence of sufficient dynamics. This highlights the interplay between data type and model choice. Overall, this work highlights the importance of representative dynamics in training data and the potential for leveraging self-supervised approaches to enhance sEMG-PR models.
\end{abstract}


\section{Introduction}

Designing a robust and reliable Surface Electromyography-based Pattern Recognition (sEMG-PR) based myoelectric control system is challenging due to a number of confounding factors, including limb position \cite{asghar_review_2022}, electrode shift \cite{young_improving_2012}, and inter-/intra-session variability \cite{tkach_study_2010, samuel_resolving_2017}, among others. Researchers are actively investigating methods to address these challenges, and increasingly, they are employing advanced Deep Learning (DL) techniques to counter their effects \cite{wu_electrode_2022, zanghieri_temporal_2020, zia_ur_rehman_multiday_2018, cote-allard_deep_2019}. 

One area where the potential of deep learning models remains underutilized is their ability to learn complex and dynamic temporal patterns. Though deep temporal classifiers such as long short-term memory networks (LSTMs) have been employed in sEMG-PR \cite{toro-ossaba_lstm_2022}, they are typically trained using static or ramp data (wherein the user gradually increases from rest to an active class of contraction \cite{scheme_training_2013}). This approach arguably fails to capture the range of temporal dynamics seen during goal-oriented device use, including transitions between classes. It is therefore reasonable that classifiers, having not been trained on dynamic data, may find it challenging to classify such data, as highlighted in our previous works \cite{tallam_puranam_raghu_analyzing_2022, raghu_decision-change_2023}. This motivates the exploration of more dynamic training data that may better reflect the behaviours seen during real-time use of myoelectric control. 

Additionally, Scheme \textit{et al.} \cite{scheme_training_2013} saw a benefit to using ramp data despite using a simple Linear Discriminant Analysis (LDA) classifier in their work. However, as LDA is a static model with no implicit sequential modeling, it cannot truly exploit the temporal information in more dynamic training data. It is therefore reasonable to hypothesize that leveraging a combination of more dynamic training data and temporal models may potentially lead to improved classifier performance in the context of continuous myoelectric classification.

Training classifiers for myoelectric control with continuous dynamic data containing transitions seems enticing, but until recently, the user-burden associated with recording a comprehensive set of class transitions for offline training has precluded training with them. Advances in deep transfer learning and domain adaptation, however, may now offer a way to overcome this obstacle \cite{cote-allard_deep_2019, wu_transfer_2023,campbell_cross_user}, motivating the proactive exploration of the benefits of using them during training.

One major complication that arises when training with continuous dynamic data is the challenge of unambiguously labeling data that contains transitions between a finite set of motion classes. These transitions create ambiguity as it becomes difficult to precisely pinpoint where one class ends and another begins. Furthermore, the classification of a subset of contractions within the larger set of those possible by the musculoskeletal system necessarily introduces undefined regions that must be traversed when transitioning from one class to another. Traditional supervised learning approaches often rely on clear, discrete labels for each data point, which can be problematic when dealing with such continuous and dynamic movements.

A potential approach to addressing this labeling challenge is to draw on recent advancements in Self-Supervised Learning (SSL) methods \cite{balestriero_cookbook_2023}. These methods leverage the inherent structure within the data itself to learn meaningful representations, eliminating the need for explicit labels. By formulating specific learning objectives or `pretexts', SSL models can capture the essential characteristics of the data in a latent space, offering a promising solution for training on unlabeled, continuous dynamic data.

While our previous conference work \cite{raghu_enabling_2024} briefly introduced the potential benefits of training with continuous dynamic data for sEMG classification, this study offers a more comprehensive and in-depth analysis. In particular, we expand upon the preliminary findings by 1) providing a detailed analysis of the impact of continuous dynamic data on classifier performance, 2) conducting a more in-depth analysis of the interplay between classifier type and training data type, including an evaluation of SSL approaches, 3) demonstrating that our initial findings, previously obtained with a Gated Recurrent Unit (GRU)+Nearest Centroid approach, generalize to a distinct end-to-end deep architecture utilizing an LSTM+linear classification head, and 4) investigating the impact of confidence-based rejection on classifier performance, particularly highlighting the effectiveness of the SSL approach in this context.

\section{Background}

\subsection{Deep Classifiers}
Deep Learning (DL) models have emerged as a powerful tool in recent years for sEMG-PR, often outperforming conventional machine learning approaches \cite{wu_electrode_2022, zanghieri_temporal_2020, zia_ur_rehman_multiday_2018, cote-allard_deep_2019, toro-ossaba_lstm_2022}. This superior performance can be attributed to DL's capacity to learn complex representations from data, capturing intricate patterns and relationships that may be difficult for traditional methods to discern. A more extensive discussion of deep learning techniques in sEMG-PR can be found in \cite{xiong_deep_2021}.

Convolutional Neural Networks, for instance, are excellent at leveraging spatial information, and thus may learn inter-electrode relationships during contractions. This makes them an excellent choice particularly when classifying high-density data, as shown by \cite{tam_fully_2020, chen_hand_2021}. However, sEMG signals also exhibit temporal dynamics that hold valuable information about user intent. These dynamics might encode, for example, the onset of a transition between different classes or changes in contraction intensity within a single class.

While LSTMs and other deep temporally aware models are adept at leveraging temporal dependencies in sequential data, their full potential in real-time myoelectric control tasks is yet to be fully explored. Existing research using deep temporal models for sEMG-PR has primarily focused on \emph{discrete gesture recognition} tasks \cite{toro-ossaba_lstm_2022, li_intelligent_2020, simao_emg-based_2019,eddy_bigdata}. Some studies have begun to explore real-time continuous myoelectric control with architectures that capture temporal information, such as the work by Betthauser et al.  \cite{betthauser_stable_2020}. In their work, the researchers trained a Temporal Convolutional Network with continuous dynamic data. Their approach relied on a labeling strategy that assumed a clear inflection point between classes, identified based on specific hand positions. While their work demonstrates the potential of temporal models for achieving higher accuracy and stability compared to non-temporal approaches, classification accuracies reported for both approaches reported in their study were below accuracy values reported in the broader literature \cite{zia_ur_rehman_stacked_2018}. This suggests that there is still room for improvement in leveraging the dynamics of myoelectric signals.

\subsection{Self-Supervised Learning}

\begin{figure*}[htbp]
     \centering
         \includegraphics[width=\textwidth]{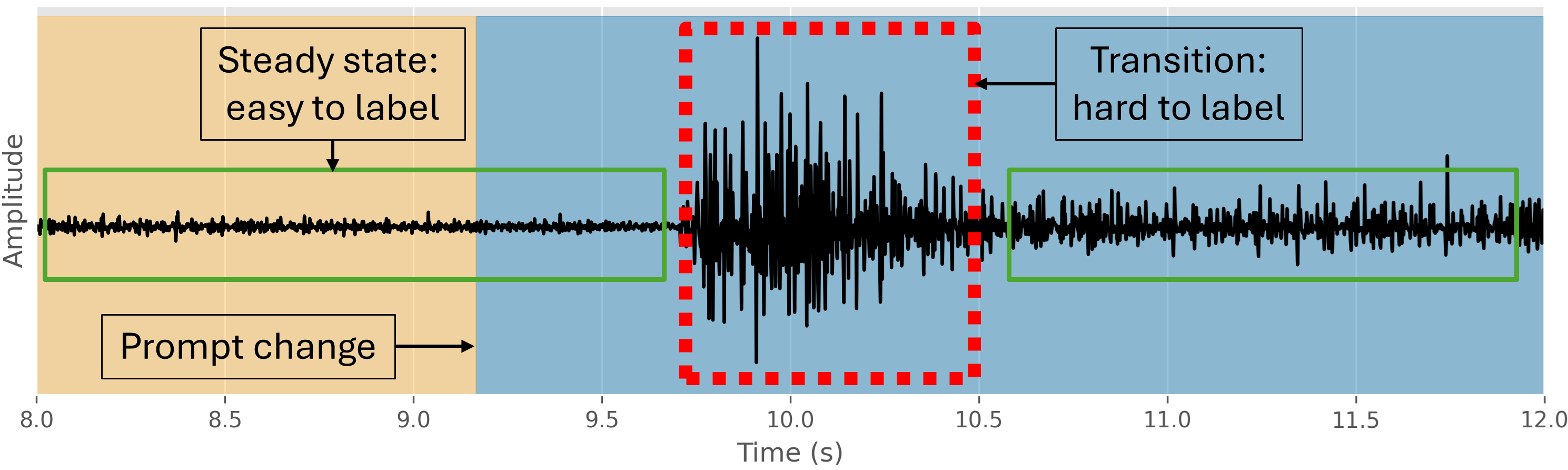}
     \caption{This figure presents an sEMG time series during an example transition from one contraction to another, illustrating the difficulty in labeling transitions (red dotted box) compared to easily identifiable steady-state regions (green boxes). The ambiguity in defining transitions, whether as inflection points or zones of uncertainty, poses a challenge for fully supervised learning approaches.}
     \label{fig:labeling_issue}
 \end{figure*}

Conventionally, most deep classifiers are trained with cross-entropy (XEnt) loss, therefore, requiring labeled data. Labeling data collected from static contractions is generally straight forward because, for offline training, they are recorded as steady-state segments of a prompted class and are void of transitions.  Ramp data are also easy to label even though they encompass a transition from no motion to steady-state by design, because the bound between no motion and steady-state in each segment can be easily identified through amplitude thresholding \cite{scheme_training_2013}. 

Continuous dynamic data is challenging to label because the bounds between steady-state and transition regions are not easily identifiable, as illustrated in Figure~\ref{fig:labeling_issue}. Furthermore, there is ambiguity in assigning labels to transition regions: should they be classified as one of the existing contraction classes or as a distinct `unknown' class, as we proposed in our previous work \cite{raghu_enabling_2024}? This lack of unequivocal labeling can lead to detrimental consequences, as deep learning models are known for their ability to memorize even random associations \cite{zhang_understanding_2021}, potentially hindering their ability to learn meaningful representations and generalize to real-world scenarios.

\begin{figure*}[htbp]
    \centering
        \includegraphics[width=\textwidth]{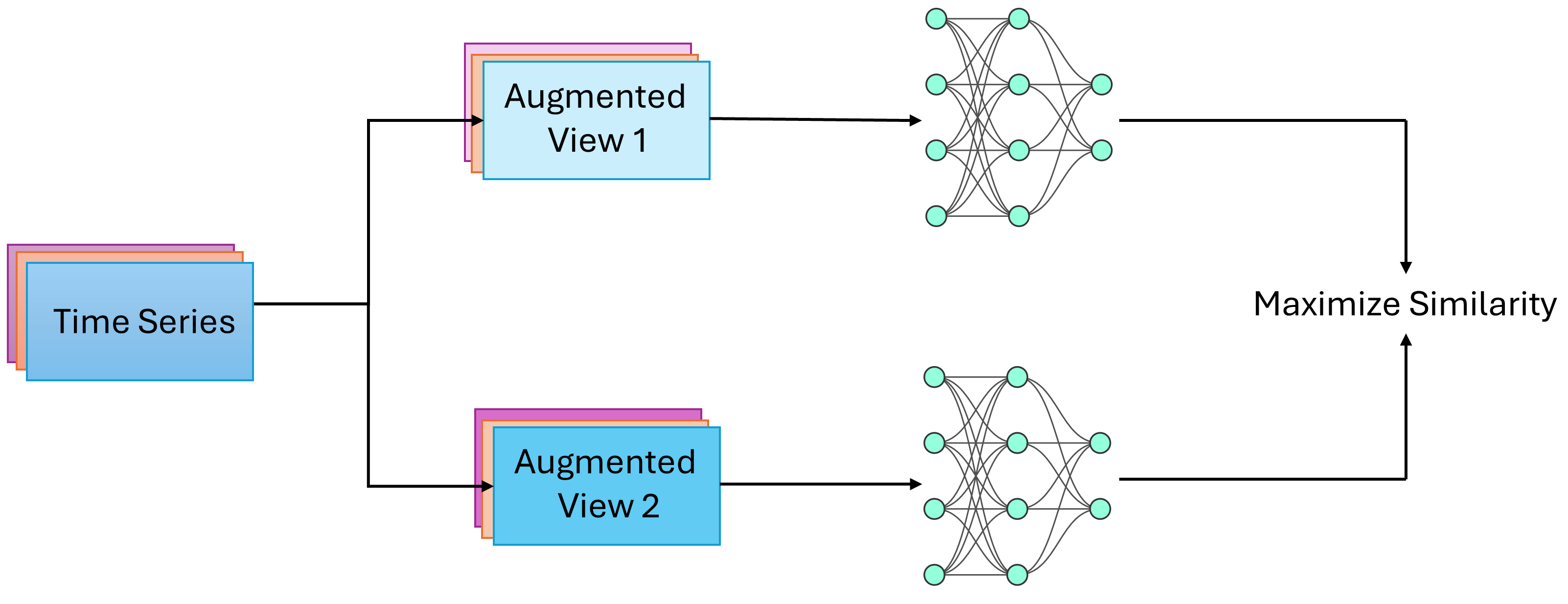}
    \caption{Illustrative example of an invariance-based SSL approach: Two augmented views are generated from the input batch and passed through the network to obtain their respective representations. The loss function then optimizes the model to produce similar representations for these augmented views, encouraging it to learn features that are invariant to the applied transformations. Note that in this work, the networks in the two branches share the same weights.}
    \label{fig:invariance_ssl}
\end{figure*}

To overcome the limitations of traditional supervised learning in the face of hard-to-label or unlabeled data, SSL has emerged as a powerful technique across various domains. SSL achieves this by constructing pretext tasks that allow the model to learn the inherent structure and relationships within the data without explicit labels. SSL methods can be broadly categorized into two main groups depending on their pretext \cite{assran_2023_CVPR}:

\begin{itemize}
    \item \textbf{Invariance-based methods}: These methods encourage the model to learn representations that are invariant or robust to various transformations or augmentations of the input data, as illustrated in Figure~\ref{fig:invariance_ssl}. For example, in SimCLR \cite{chen_simple_2020}, different augmented views of the \emph{same} input data are treated as positive pairs, and the model is trained to maximize their similarity while minimizing similarity to other images. This encourages the model to learn representations that capture the core essence of the data and disregard noise. Variance-Invariance-Covariance Regularization (VICReg) \cite{bardes_vicreg:_2022}, the method employed in this study, also falls under this category, utilizing a combination of variance, invariance, and covariance regularization to learn meaningful representations without relying on explicit pairwise comparisons.
    \item \textbf{Generative methods}: These methods involve training the model to generate or reconstruct the original data from a compressed or latent representation. By learning to generate realistic outputs, the model implicitly captures the underlying structure and patterns within the data. Examples of generative SSL methods include next-step prediction \cite{liu_self-supervised_2023} and Masked Autoencoders  \cite{he_2022_CVPR}.
\end{itemize}

A comprehensive overview of the different SSL approaches, including discussions on the critical role of augmentations and the challenges of representation collapse, is provided in\cite{balestriero_cookbook_2023}. The representations learned during SSL can be effectively transferred to various `downstream tasks' --- tasks that utilize the pre-trained representations for a specific objective, such as time series classification ---  by adding a simple classification head on top of the pre-trained SSL model. The efficacy of SSL has spurred investigations into their application in diverse domains including natural language processing \cite{jian_non-linguistic_2022}, time-series analysis \cite{yang_timeclr:_2022}, and medical image analysis \cite{azizi_big_2021}.

To further enhance the effectiveness of SSL, data augmentations such as random cropping, flipping, and adding noise are commonly applied to the training data during pretext training. These artificially expand the training data by creating variations of existing data points. This increased data diversity exposes the model to a wider range of feature values, ultimately improving its ability to generalize to unseen data during inference. Additionally, data augmentation helps prevent overfitting by forcing the model to learn from variations of the same data, preventing it from memorizing specific characteristics of the unlabeled data itself. Finally, certain types of augmentations, like scaling, can encourage the model to learn representations that are invariant to irrelevant perturbations in the original feature space, such as the amplitude of the signal.  These concepts are evocative of many of the current confounds in myoelectric control, wherein models degrade when exposed to unseen limb positions, electrodes shifts, or control dynamics, further motivating their exploration in this work \cite{asghar_review_2022}.  

\section{Methods}
To evaluate the impact of leveraging continuous dynamic data in offline training, we compared the performance of classifiers trained on ramp data versus those trained on continuous dynamic data. We employed the commonly used LDA classifier as delineated in \cite{scheme_confidence-based_2013} as a benchmark. We also employed a deep LSTM classifier based on the hypothesis that it could more effectively utilize the temporal aspects of the training data.  The specific choice of models was not the main focus of this work, but instead the ability (or lack thereof) to model temporal dynamics. Both classifiers underwent supervised training with each of the sets of data outlined below, for a complete comparison. Supervised training with continuous dynamic data was accomplished using a naive labeling strategy, also explained below. An additional LSTM classifier was trained with the continuous dynamic data via self-supervision.  Details about the data sets, classifiers, labeling strategies, and training and testing protocols are provided below. The software stack used for the study was: 

\begin{itemize}
    \item \textbf{Processing}: Numpy (v1.26) \cite{harris_array_2020}, SciPy (v1.12)\cite{virtanen_scipy_2020}, 
    \item \textbf{Pattern Recognition}: Scikit-learn (v1.4) \cite{pedregosa_scikit-learn:_2011} , Keras \cite{chollet2015keras} / TensorFlow (v2.15) \cite{tensorflow2015-whitepaper}, 
    \item \textbf{Statistical Analysis and Visualization}: Pandas (v2.2) \cite{mckinney_data_2010}, Matplotlib (v3.8) \cite{hunter_matplotlib:_2007} / Seaborn (v0.13) \cite{waskom_seaborn:_2021}, Pingouin (v0.5) \cite{vallat_pingouin:_2018}.
\end{itemize}

\subsection{Training and Testing Dataset}

A previously collected dataset, described in \cite{raghu_decision-change_2023}, was used in this work. The study was approved by the University of New Brunswick’s Research Ethics Board (REB \#2021-116). In short, the dataset consists of sEMG data recorded from forty-three able-bodied participants (age: $25.98 \pm 5.8$, 26 male, 17 female) using six Delsys Trigno \textregistered bipolar electrodes placed around the circumference of the forearm, a third of the way down, distal to the elbow. The participants were recruited mainly from a graduate student population and gave informed consent. The electrodes were affixed in an equidistant clockwise fashion starting above the middle of the flexor carpi radialis muscles. The acquisition device sampled the sEMG signals at \SI{2}{\kilo\hertz} and digitized the signals with a 16-bit Analog-To-Digital converter. A Leap Motion Controller (LMC) was used to capture hand position data as participants performed prompted contractions, providing a reliable ground truth for identifying transition regions in the dynamic trials.

Each participant completed 5 ramp trials and 6 continuous dynamic trials. All trials were guided by a visual computer prompt and comprised the following contractions: Wrist Flexion (WF), Wrist Extension (WE), Wrist Pronation (WP), Wrist Supination (WS), Hand Close (HC), and Hand Open (HO). No Movement (NM) was also included yielding 7 classes in total.  Each ramp trial included a set of 6 ramp contractions. Participants were shown a prompt for the target contraction and then given \SI{3}{\second} to gradually ramp up their muscle activity from a neutral position to a steady-state contraction. Following this, they were given \SI{3}{\second} to relax and return to a neutral position while being shown the prompt for the next contraction. Data collection occurred only during the \SI{3}{\second} ramp-up phase; the relaxation periods were not recorded. The total ramp data amounted to 5 trials x 7 prompts / trial x \SI{3}{\second} / prompts = \SI{105}{\second} per participant.

Each continuous dynamic trial consisted of 42 transitions, encompassing all possible combinations between the 7 classes, resulting in 43 steady-state regions. Participants were presented with a visual prompt for \SI{3}{\second} seconds, indicating the target contraction. They were then expected to transition from their current state to the prompted contraction in a smooth and natural manner, and maintain the new contraction until the next prompt appeared. Data collection occurred continuously throughout the trial, capturing both the steady-state contractions and the transitions between them. The LMC data was used to identify the bounds of transition regions based on the velocity of movement as calculated from the extracted hand vectors. The continuous dynamic data amounted to 6 trials x 43 prompts / trial x \SI{3}{\second} / prompts = \SI{774}{\second} per participant.

We intend to make this dataset publicly available within LibEMG \cite{eddy_libemg:_2023} in the near future.

\subsection{Data Processing and Labeling}
EMG signals were bandpass filtered \qtyrange{20}{450}{\hertz} using a \nth{4} order zero-phase filter to remove any low or high frequency noise \cite{simao_review_2019, samuel_intelligent_2019}. Each trial was then segmented into overlapping frames with a length of \SI{162}{\milli\second} and an increment of \SI{13.5}{\milli\second}. The increment value was selected to match the packet rate of the Delsys Trigno device. The Low-Sampling Frequency 4 (LSF4), consists of 4 features:  L-scale, maximum fractal length, mean square root, and Willison amplitude \cite{phinyomark_feature_2018}, features were extracted from each frame. This feature set was demonstrated to perform better than other feature sets and was resilient to low sampling rates, and thus was chosen for this work. Labels for each frame were then assigned for supervised training as follows:

The ramp data was labeled as per existing studies \cite{scheme_training_2013}. First, all frames were labeled based on their corresponding prompts.  Then a NM threshold  was established based on the amplitudes (i.e. mean absolute value) of the NM frames.  The threshold was calculated using the  mean ($\mu$) and standard deviation ($\sigma$) across NM frames according to ($\mu + 3 \times \sigma$). Finally, frames with motion class labels that had amplitudes below the threshold were relabeled as NM class.

In labeling frames for continuous dynamic data, a strategy somewhat similar to the the one used by Betthauser \textit{et al.} \cite{betthauser_stable_2020} was employed. The first frame that followed a prompt change and aligned with the onset of movement as detected by the LMC was considered the start of a transition to the next motion class (i.e the prompted class).  This frame, and all subsequent frames were then labeled according to the prompt.  This process was repeated for each prompt.  The strategy was chosen to encourage the model to transition to subsequent classes swiftly, hopefully resulting in a more responsive model. The researchers in \cite{betthauser_stable_2020} relied on hand \emph{position} to differentiate between the previous and upcoming classes, whereas we utilize \emph{velocity} information. This distinction was chosen because velocity captures the dynamics of movement initiation, which may be a more reliable indicator of class transition compared to relying solely on static hand position.

\subsection{VICReg Loss}

In this study, we explore the potential of SSL for sEMG-PR classification, employing the Variance-Invariance-Covariance Regularization (VICReg) loss function \cite{bardes_vicreg:_2022} for its simplicity and, importantly, its compatibility with small batch sizes, which are typical for sEMG-PR applications. VICReg is briefly described below. 

Let $X \in \mathbb{R}^{N \times T \times F} $ denote a batch of data containing $N$ samples each of which is a $T \times F$ time-series where $T$ is the length of the time-series and $F$ is the feature dimensionality. Let $\widetilde{X}$ and $X'$ be two augmented views of $X$. Let the deep network with trainable parameters $\theta$ be denoted by $f_{\theta}(.)$, and let the latent space embeddings produced by the network for $\widetilde{X}$ and $X'$ be denoted by $f_{\theta}(\widetilde{X}) = \widetilde{Z}$ and $f_{\theta}(X') = Z'$, where $\widetilde{Z}, Z' \in \mathbb{R}^{N \times D} $, and $D$ represents the output dimensionality. The subscripts, (e.g. $z_i$) are used to denote indexing a single sample in a tensor and superscripts (e.g., $z^j$) denote indexing a single dimension in a tensor. The VICReg loss function is:
\begin{multline}
    l(\widetilde{Z}, Z') \\ = \lambda s(\widetilde{Z}, Z') + \mu[v(\widetilde{Z}) + v(Z')] + \nu[c(\widetilde{Z}) + c(Z')]
    \label{eqn:vicreg_main}
\end{multline}
where $s(.)$ is the invariance term which encourages the model to produce similar embeddings for the two views of the data and is given by:
\begin{equation}
    s(\widetilde{Z}, Z') = \frac{1}{N} \sum_{i=1}^{N} \left\Vert\ \widetilde{z}_i - {z'}_i \right\Vert^2
    \label{eqn:vicreg_invariance}
\end{equation}
and $v(.)$ is the variance term preventing collapse of the outputs, given by:
\begin{equation}
    v(Z) = \frac{1}{D} \sum_{j=1}^{D} ReLU(\gamma - S(z^j, \epsilon))
    \label{eqn:vicreg_variance}
\end{equation}

In  \ref{eqn:vicreg_variance}, $S(.)$ is the regularized standard deviation estimated from the batch of data, given by:
\begin{equation}
    S(z, \epsilon) = \sqrt{Var(z) + \epsilon}
\end{equation}
where $\epsilon$ is a small constant used to regularize the standard deviation.
The final term in \ref{eqn:vicreg_main}, $c(.)$, is the covariance term that encourages the network to produce distinct information across the output dimensions.  It is given by:
\begin{equation}
    c(Z) = \frac{1}{D} \sum_{i, j, i \ne j}^{D} [C_{ij}]^2
\end{equation}
where $C \in \mathbb{R}^{D \times D} = Cov(Z)$ is the covariance matrix estimated from the batch of embeddings.

The terms $\lambda, \mu, \nu$ are hyperparameters that control the relative importance of the invariance, variance, and covariance terms in the loss function respectively, and $\gamma$ is a constant target standard deviation value. In this study, the these hyperparameters were set as: $\gamma = 1, \lambda = 8, \mu = 32, \nu = 1$ based on preliminary experimentation with a random $10\%$ subset of the continuous dynamic data. While a more extensive hyperparameter search might yield further improvements, these values were found to provide satisfactory performance in our initial exploration.

\subsection{Deep Temporal Network Architecture}

\begin{figure*}[htbp]
    \centering
        \includegraphics[width=1\linewidth]{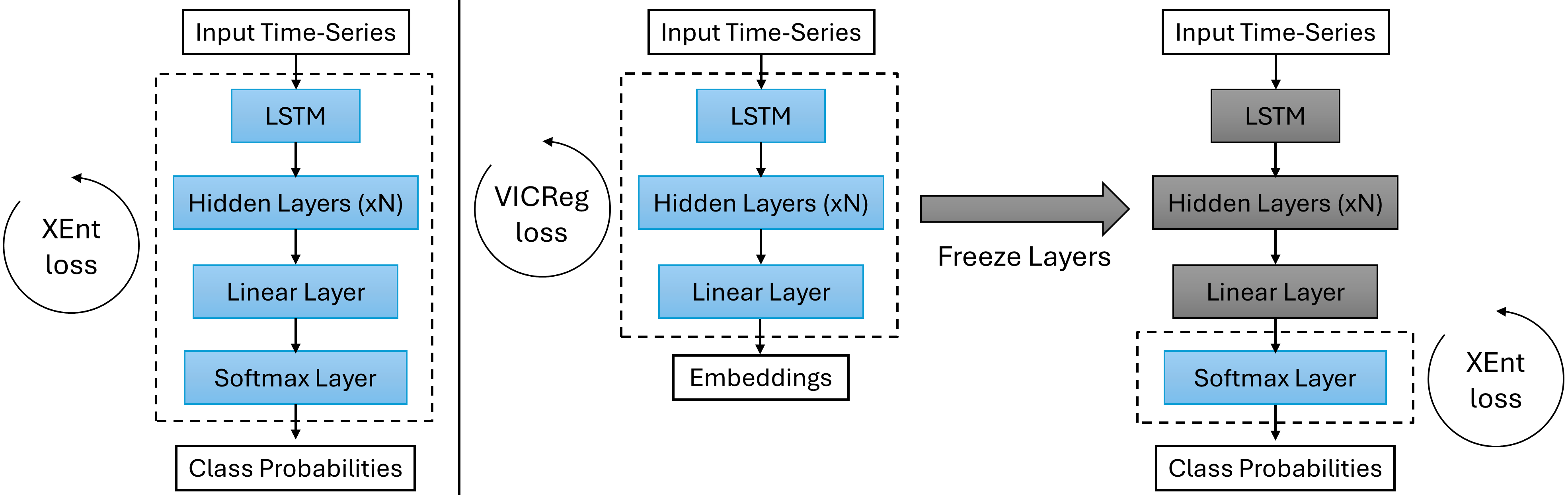}
    \caption{The deep network architecture comprises a backbone and a classification head (softmax layer). In conventional supervised training (left), both are optimized jointly using XEnt loss. In the proposed method (right), the backbone is first pre-trained with VICReg and then frozen. The classification head is trained on top using XEnt loss.}
    \label{fig:lstm_architecture}
\end{figure*}

Our temporal network, as depicted in Figure~\ref{fig:lstm_architecture}, builds upon the architecture presented in our previous work \cite{raghu_enabling_2024}, but incorporates key modifications. Like the previous model, it follows a modular design with a trainable backbone and a classification head. However, we replaced the GRU layer in the backbone with an LSTM layer, aiming to leverage its potential for capturing even more complex temporal dependencies in the sEMG signals. We also adopted a linear layer with softmax activation for the classification head, offering a more streamlined end-to-end approach.

During fully-supervised training with labels, both the backbone and the classification head were trained jointly with the XEnt loss function to optimize the overall classification performance. However, when employing SSL, we leveraged a different training strategy. Here, the focus was on learning transferable representations within the backbone. Therefore, we first trained the backbone with no labels and VICReg loss and then froze its parameters. Subsequently, the linear classification head was added on top of the frozen backbone, and only the parameters of the classification head were fine-tuned during training with labeled data and XEnt loss. This approach allowed the pre-trained backbone to act as a dynamics-informed feature extractor, while the classification head learned to map these features to specific class labels.

In this study, the backbone featured an LSTM layer with $128$ units followed by $2$ hidden layers. Each hidden layer consisted of a Dense layer containing $128$ units followed by layer normalization and ReLU activation. The outputs were then finally routed through a dense layer with linear activation.
The classification head consisted of a Dense layer with $7$ units (equal to the number of classes) with softmax activation. These hyperparameter values were based on empirical observations and prior experience, and were not extensively optimized for this specific dataset.

\textit{Training Details}: AdamW \cite{loshchilov_decoupled_2019} was used as the optimizer in all cases. The batch size was set to $256$. The learning rate was set to $\SI{1e-3}{}$ when training the backbone and classification head separately, and was set to $\SI{1e-4}{}$ when training end-to-end with XEnt loss. In all cases, early stopping was employed to stop the training after $10$ epochs of no improvement in validation loss to prevent overfitting. All features were standardized to have zero mean and unit standard deviation based on the training data in all cases.

\subsection{Augmentations}

Augmentations play a crucial role in self-supervised learning. While computer vision tasks have seen numerous effective augmentations proposed in the literature, time-series tasks have garnered less focus. Researchers \cite{iwana_empirical_2021} have surveyed several potential augmentation strategies for time-series, recognizing the task's complexity due to the diverse characteristics of time-series data across different domains. To the best of our knowledge, there appears to be no established research on effective augmentations for sEMG-PR applications. For this work, we adopted 3 simple augmentations: random lag, amplitude scaling, and additive white Gaussian noise (AWGN) as described below. 

Let $x_i \in \mathbb{R}^{T \times F} $ be a sample time-series. Such a sample can be represented as a matrix with rows corresponding to time steps $t = t_1 \ldots t_T$ and columns corresponding to features. For simplicity, we assume $t_i$ to be integer indices.
\begin{equation}
    x_i = \begin{bmatrix}
    x_{t_1 1} & x_{t_1 2} & \ldots & x_{t_1 F}\\
    x_{t_2 1} & x_{t_2 2} & \ldots & x_{t_2 F}\\
    \vdots & \vdots & \vdots & \vdots\\
    x_{t_T 1} & x_{t_T 2} & \ldots & x_{t_T F}\\
\end{bmatrix}
\end{equation}

\subsubsection{Random lag}
Let $\phi$ be a random variable following the discrete uniform distribution over the set $\{a, a+1, \ldots, b\}$ where $b, a$ are integers, and $b \ge a$. Random lag is defined as adding a random offset $\phi$ to all the time steps:
\begin{equation}
    \widetilde{x_i} = \begin{bmatrix}
    x_{(t_1 + \phi) 1} & x_{(t_1 + \phi) 2} & \ldots & x_{(t_1 + \phi) F}\\
    x_{(t_2 + \phi) 1} & x_{(t_2 + \phi) 2} & \ldots & x_{(t_2 + \phi) F}\\
    \vdots & \vdots & \vdots & \vdots\\
    x_{(t_T + \phi) 1} & x_{(t_T + \phi) 2} & \ldots & x_{(t_T + \phi) F}\\
\end{bmatrix}
\label{eqn:aug_lag}
\end{equation}

Random lag augmentation introduces a variation where a portion of the time series is shifted. This approach can be seen as an alternative to the windowing strategy in \cite{iwana_empirical_2021}, where both the original and shifted segments are derived from a larger sequence. By learning from these temporally adjacent yet shifted versions, the network is encouraged to recognize the inherent similarity between these points and position them closer together in the latent space. This, in turn, promotes a focus on the overall structure of the time series, regardless of minor temporal shifts.

\subsubsection{Random Scaling}
Let $\mathrm{A} \in \mathbb{R}^{F}$ be a vector $
\mathrm{A} = \begin{bmatrix} \alpha_1 & \alpha_2 & \ldots \alpha_F\\
\end{bmatrix}$ whose elements $\alpha_i$ are drawn from a normal distribution, $\alpha_i \sim \mathcal{N}(\mu,\,\sigma^{2})$. Random scaling is defined as multiplying a random scaling factor to all the time steps for a given feature:
\vspace{-1mm}
\begin{align}
    \widetilde{x_i} &= \mathrm{A} \circ x_i    
\label{eqn:aug_scale}\\[10pt]
    &= \begin{bmatrix}
    \alpha_{1}x_{t_1 1} & \alpha_{2}x_{t_1 2} & \ldots & \alpha_{F}x_{t_1 F}\\
    \alpha_{1}x_{t_2 1} & \alpha_{2}x_{t_2 2} & \ldots & \alpha_{F}x_{t_2 F}\\
    \vdots & \vdots & \vdots & \vdots\\
    \alpha_{1}x_{t_T 1} & \alpha_{2}x_{t_T 2} & \ldots & \alpha_{F}x_{t_T F}\nonumber\\
\end{bmatrix}
\end{align}

In Equation \ref{eqn:aug_scale}, the symbol $\circ$ denotes the Hadamard (element-wise) Product. This augmentation promotes network resilience to feature scale variations, which are anticipated, for example, during changes in contraction intensity. 

\subsubsection{Additive White Gaussian Noise}

AWGN is a simple and popular augmentation where each sample is corrupted with additive noise. Let $\Xi \in \mathbb{R}^{T \times F} $ be a matrix whose elements $\xi_{ij}$ are drawn from a normal distribution, $\xi_{ij} \sim \mathcal{N}(\mu,\,\sigma^{2})$. The augmentation is given by
\vspace{-1mm}
\begin{align}
    \widetilde{x_i} &= x_i + \Xi
\label{eqn:aug_awgn}\\[10pt]
    &= \begin{bmatrix}
    x_{t_1 1}  + \xi_{11} & x_{t_1 2} + \xi_{12} & \ldots & x_{t_1 F} + \xi_{1F}\\
    x_{t_2 1}  + \xi_{21} & x_{t_2 2} + \xi_{22} & \ldots & x_{t_2 F} + \xi_{2F}\\
    \vdots & \vdots & \vdots & \vdots\\
    x_{t_T 1}  + \xi_{T1} & x_{t_T 2} + \xi_{T2} & \ldots & x_{t_T F} + \xi_{TF}\nonumber
\end{bmatrix}
\end{align}

This augmentation enhances the network's resilience to noise within the features. Given that sEMG signals are inherently noisy, the features derived from these signals are likely to be noisy as well. Consequently, bolstering the network's ability to withstand such noise could enhance model performance.

\begin{figure}[htb!]
    \centering
        \includegraphics[width=0.5\linewidth]{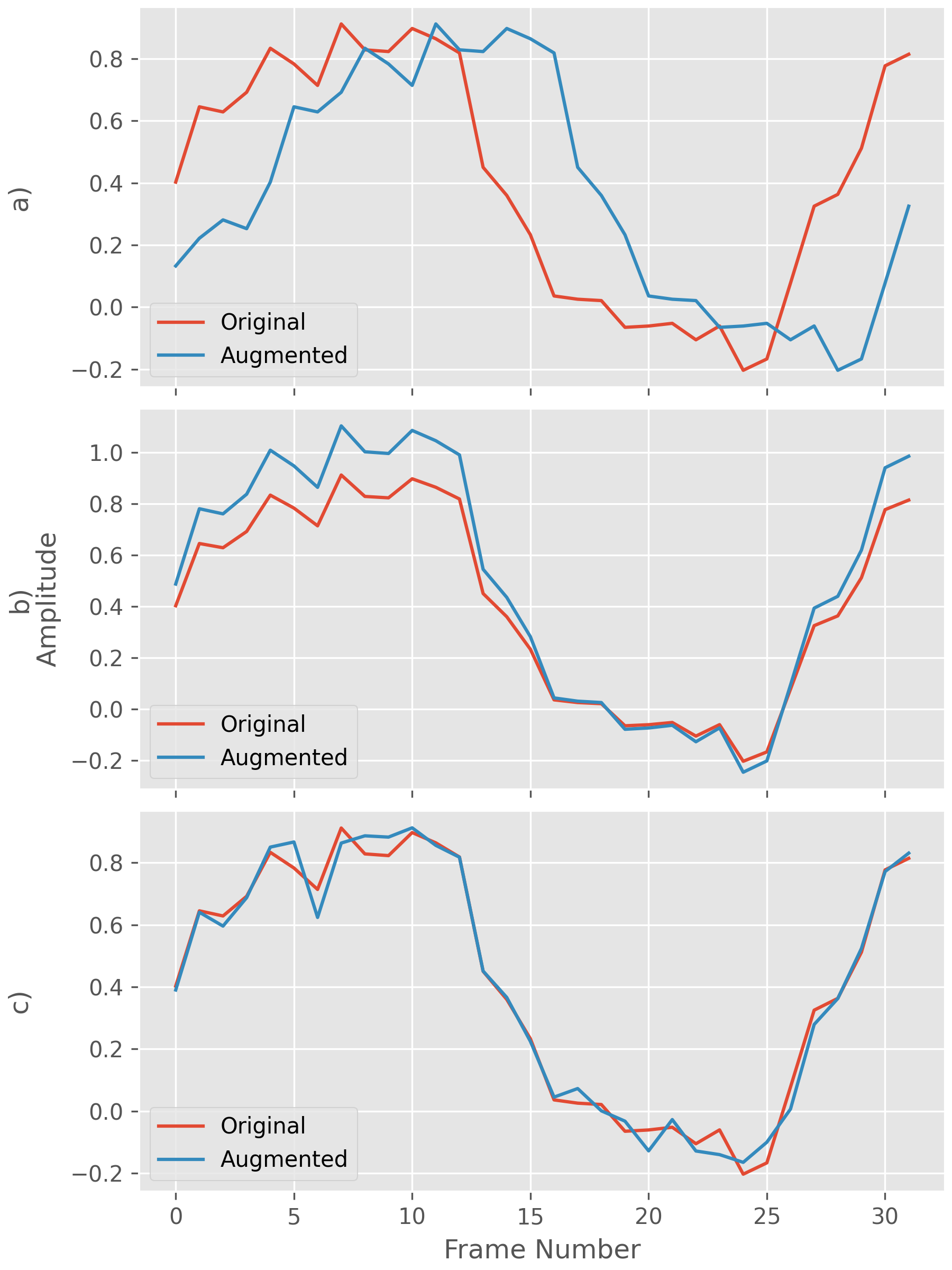}
    \caption{Time-series example with individual augmentations: (a) Random Lag, (b) Random Scaling, (c) AWGN.}
    \label{fig:time_series_augmentation}
\end{figure}

The augmentation pipeline in this study consisted of three sequential operations: 1) random lag ($a=-4, b=4$), 2) random scaling ($\mu = 1, \sigma = 0.05$), and 3) corruption through AWGN ( $\mu = 0, \sigma = 0.05$). All hyperparameters were based on preliminary experimentation with a random $10\%$ subset of the continuous dynamic data. An example of a feature time-series, augmented by each augmentation method, is illustrated in the Figure \ref{fig:time_series_augmentation}.

\subsection{Evaluating Model Performance}

The five classifiers evaluated in this work were: LDA trained on labeled ramp data (LDA-R), LSTM trained on labeled ramp data (LSTM-R), LDA trained on labeled continuous dynamic data (LDA-D), LSTM trained on labeled continuous dynamic data (LSTM-D), and LSTM pre-trained on unlabeled continuous dynamic data with VICReg loss (LSTM-V). LDA does not allow for this final kind of training, thus there was no LDA-V.  

All classifiers were evaluated on continuous dynamic data using 7 of the offline continuous transition performance metrics proposed as part of our previous work \cite{tallam_puranam_raghu_analyzing_2022}. These metrics provide a comprehensive picture of classification performance as they capture errors during steady-state and transitions, as well as responsiveness of the classifiers during transitions between classes.

\begin{enumerate}
    \item \textbf{Steady-state Active Error Rate (SS-AER)}: The percentage of classifier decisions in steady-state that differ from the ground truth, excluding decisions that were no movement. Lower values are better. This metric captures the overall accuracy of the classifier, but excludes the NM class to reflect its lower impact (i.e., doing nothing is better than making an incorrect movement \cite{robertson_effects_2019}).
    \item \textbf{Steady-state Total Error Rate (SS-TER)}: The percentage of classifier decisions in steady-state that differ from the ground truth, \emph{including} when the decisions were no movement. Lower values are better. This metric complements AER and captures the overall accuracy of the classifier. It is possible to artificially reduce AER by always outputting NM class, but this would increase TER.
    \item \textbf{Steady-state Instability (SS-INS)}: The percentage of consecutive decisions that differ from one another, excluding blips to NM class. Lower values are better. This metric captures spurious fluctuations or blips that occur in the decision stream when holding a contraction in steady-state.
    \item \textbf{Offset delay (TOFF)}: The number of frames between movement initiation in response to a prompt change to when the classifier leaves the previous steady-state. Lower values are better. This metric captures the responsiveness of the classifier when switching from one class to another.    
    \item \textbf{Transition duration (TTD)}: The average number of frames a transition lasts. Lower values are better.
    \item \textbf{Transition Instability (INS)}: The average number of consecutive decisions that differ from one another in the transition region, excluding blips to NM class. Lower values are better. Similar to SS-INS, this metric captures spurious changes in the decision stream that occur during a transition.
    \item \textbf{Tertiary Class Errors (TCE)}: The average number of decisions in the transition region that are not one of: the previous prompt, the new prompt, or no movement class. Lower values are better. This metric captures the errors during transition to other classes that are unrelated to the context of the transition.    
\end{enumerate}

For LDA-R, all five trials of the ramp data were utilized to train the LDA classifier, which was then assessed using all six trials of the continuous dynamic data. For LSTM-R, one random ramp trial was selected for validation, while the other four trials were used for training. This was necessary to promote generalization, but notably resulted in 20\% smaller training sets for the LSTM-R. The LSTM-R was assessed using the same six trials of the continuous dynamic data. 

In the cases of training with continuous dynamic data, a leave-one-trial-out method was employed for evaluation. Each participant had one trial removed for evaluation, and the remaining five trials were used to train the LDA-D. Similarly, for LSTM-D, one trial was reserved for validation, and the remaining four were used for training. This procedure was iterated six times, allowing the trained models to be evaluated on all six trials of the continuous dynamic data. The presented results represent the average performance across these 6 trials.

Analysis focused on answering the following three questions using a two-factor repeated measures ANOVA ($\alpha = 0.05$). 

\begin{enumerate}
    \item \textbf{Q1}: Is there a difference between ramp and dynamic training? (Comparing models trained on labeled ramp data (LDA-R, LSTM-R) vs. models trained on labeled continuous dynamic data (LDA-D, LSTM-D))
    \item \textbf{Q2}: Is there a difference between temporal and non-temporal models? (Comparing LSTM vs. LDA within each training data category)
    \item \textbf{Q3}: Does the LSTM-V approach do better than the rest? (Evaluating the performance of VICReg pre-trained LSTM-V against all other models)
\end{enumerate}

The fully supervised, labeled models were included in the two-factor ANOVA, with training data type and model type as factors. The Greenhouse-Geisser correction was applied to account for potential violations of sphericity, and the corresponding corrected p-values are reported.

For metrics exhibiting statistically significant effects ($p < 0.05$) and interactions, Šidák-corrected pairwise comparisons were conducted. Cohen's d effect sizes were calculated and categorized according to \cite{sawilowsky_new_2009}: $d = 0.01$ = Very Small, $d = 0.2$ = Small, $d = 0.5$ = Medium, $d = 0.8$ = Large, $d = 1.2$ = Very Large, and $d = 2.0$ = Huge. Pairwise comparisons were also employed to compare the performance of LSTM-V to the other models.

\subsubsection{Rejection Analysis}
In addition to the primary analysis, we investigated the impact of confidence-based rejection, a post-processing scheme shown to improve online usability \cite{scheme_confidence-based_2013}, on classifier performance. 

A rejection threshold was determined by examining the relationship between rejection threshold and average error rate, selecting a threshold that minimized over-rejection. Subsequently, all decisions with confidence values lower than the threshold were rejected to the NM class. Each classifier's performance was re-evaluated with and without rejection, effectively adding another layer of analysis to the existing factors (training data type and model type). While we did not conduct a formal three-factor ANOVA, this approach allowed us to assess the impact of rejection on the various performance metrics within the context of our original research questions.

\section{Results}

\begin{figure*}[htb]
    \centering        
    \includegraphics[width=\textwidth]{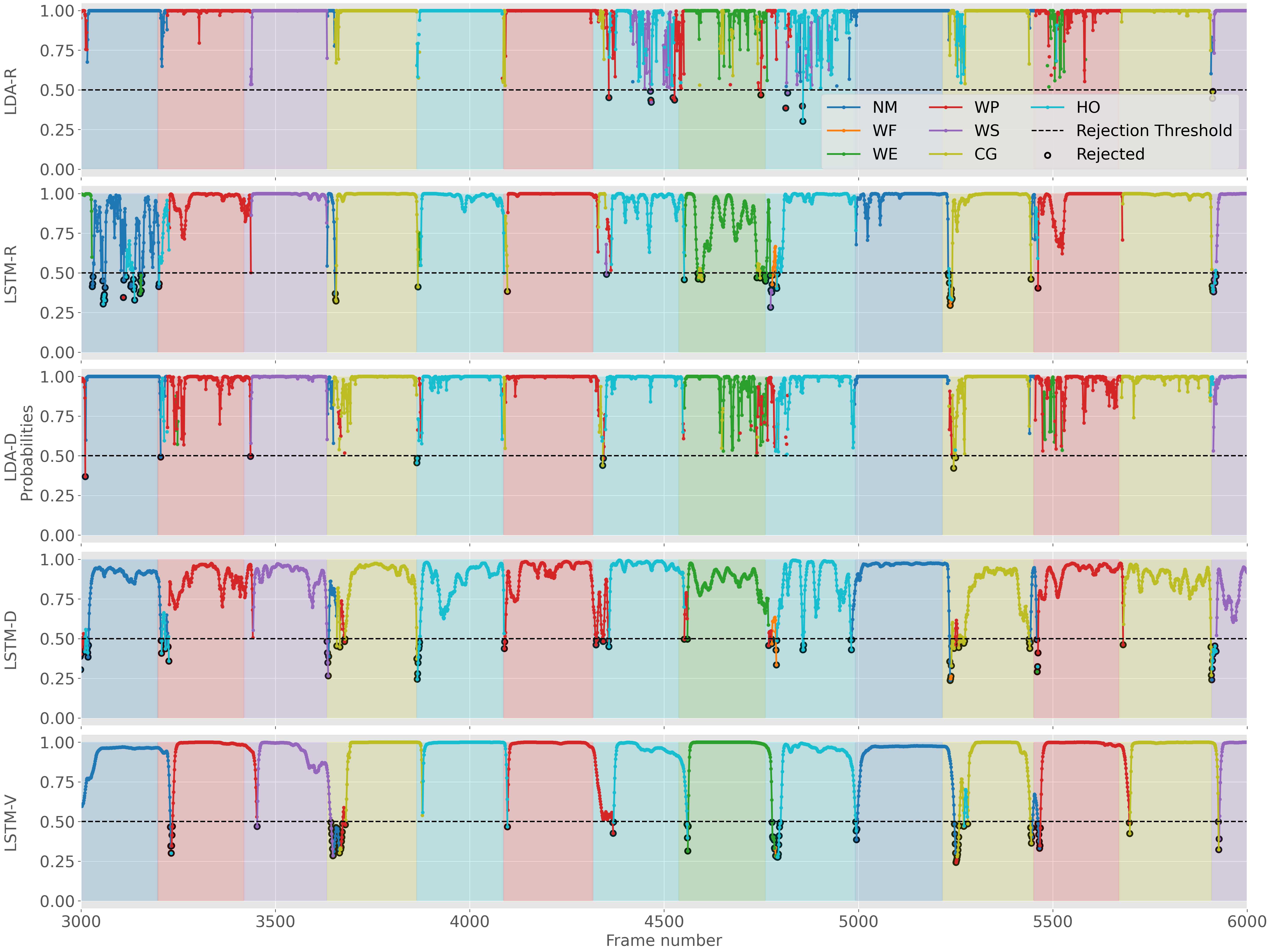} 
    \caption{Decision streams obtained from different classifiers for an example trial. Colours of the data points represent the classifiers decision (as indicated in the legend), and values represent confidence; shaded regions denote the ground truth class (based on the Leap data). A rejection threshold of $0.5$ is shown to illustrate its potential benefits across different models.}
    \label{fig:decision_stream}
\end{figure*}

\begin{figure*}[htb]
    \centering    
        \includegraphics[width=\textwidth]{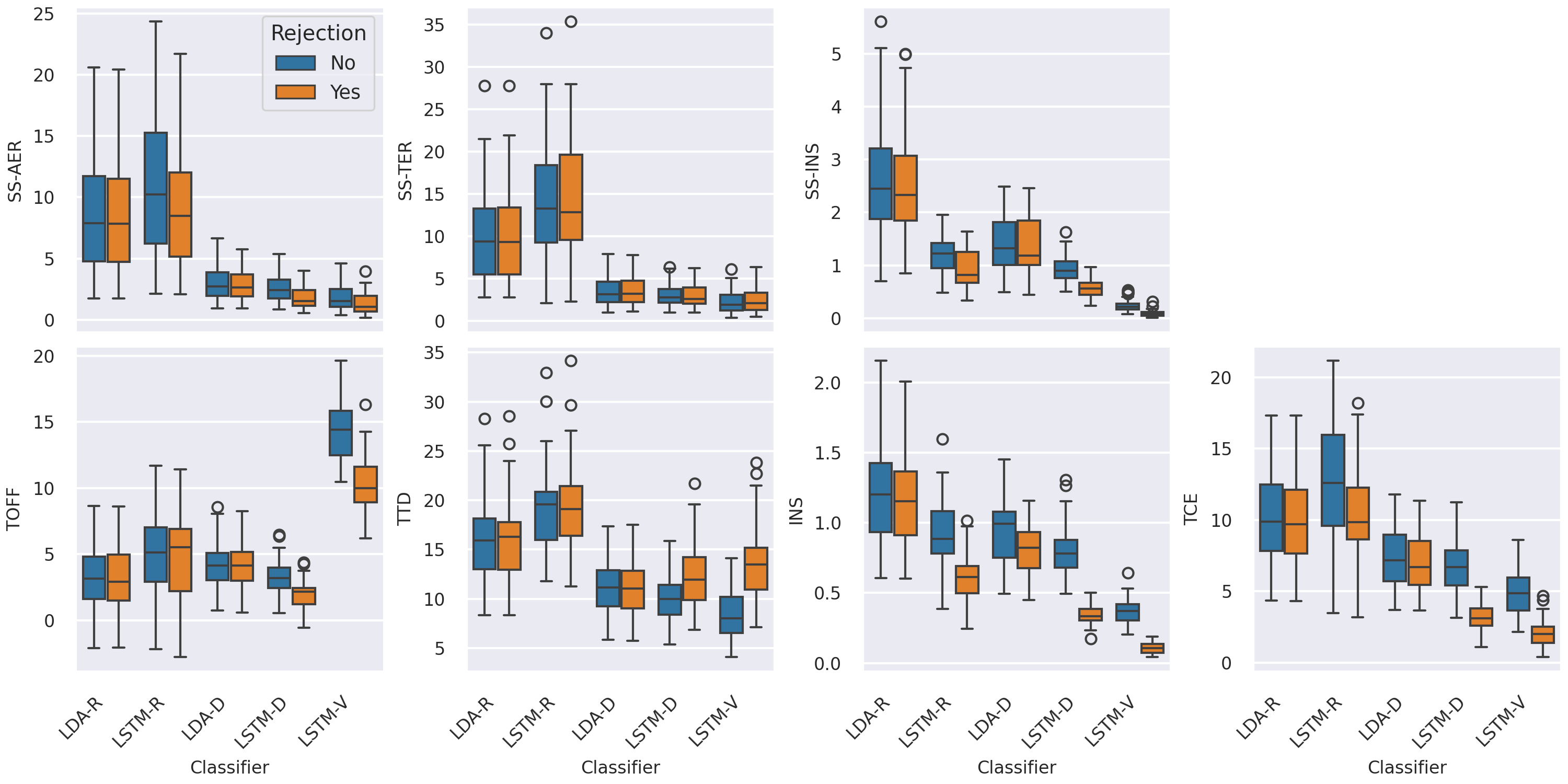}
    \caption{Comparison of the distribution of performances of the different classifiers across subjects and trials, with and without a rejection threshold of $0.5$. The top row shows steady-state metrics, whereas the bottom row shows transition metrics. Classifier schemes denoted as -R were trained with Ramp data, -D were trained with continuous dynamic data, and -V denotes the VICReg pre-trained model with the continuous dynamic data.}
    \label{fig:metrics}
\end{figure*}

\subsection*{Impact of Training Data Type and Model Type (without Rejection)}

Figure \ref{fig:decision_stream} depicts portions of the decision streams produced by the different models for a representative test trial as a way to provide visual context for the steady-state and transition performance metrics. For instance, LSTM-V appears to exhibit less TER, but through careful inspection, and increased TOFF is also apparent. This is confirmed in Figure \ref{fig:metrics}, which presents a boxplot comparing each metric averaged across trials per subject and then subjects for each model. 

Results of the statistical analysis revealed significant main and interaction effects for both training data type ($p < 0.0001$ in all cases) and model type ($p < 0.005$ in all cases) across SS-AER, SS-TER, SS-INS, TTD, INS, and TCE. An interaction effect ($p < 0.0001$) was noted in TOFF, but no main effects ($p > 0.1$ for both model type and training data type). The interaction effect between training data type and model type indicates that one factor depends on the other. 

To delve deeper into the interaction effect observed between training data type and model type, we conducted post-hoc pairwise comparisons for all the metrics. Compared to LDA-R, LDA-D achieved significantly better performance with Large or greater effect size ($p < 0.0001$; $d > 0.98$ in all cases) for all of the metrics except for TOFF, where LDA-R had statistically lower TOFF than LDA-D with a Medium effect size ($p < 0.005$; $d > 0.58$). Similarly, compared to LSTM-R, LSTM-D achieved significantly better performance with Medium effect size or greater ($p < 0.0001$, $d > 0.5$ in all cases) for all metrics except INS where no significant difference emerged. These comparisons confirmed the advantage of training with continuous dynamic data.

A more intriguing finding emerged when examining the performance within each training data type. LDA-R outperformed LSTM-R with Small or greater effect size ($p < 0.005$; $d > 0.4$ in all cases) for all the tested metrics except for INS and SS-INS where the LSTM-R was better with a Large or better effect size ($p < 0.005$; $d > 0.9$ in both cases). Conversely, LSTM-D significantly outperformed LDA-D for all tested metrics with at least a Small effect size ($p < 0.005$ ; $d > 0.35$) except for TCE where no significant differences emerged. These findings suggest that when training with continuous dynamic data, employing temporal models appears to help, but, when training with ramp data, they may decrease performance.

\begin{figure*}[htpb]
    \centering    
        \includegraphics[width=\textwidth]{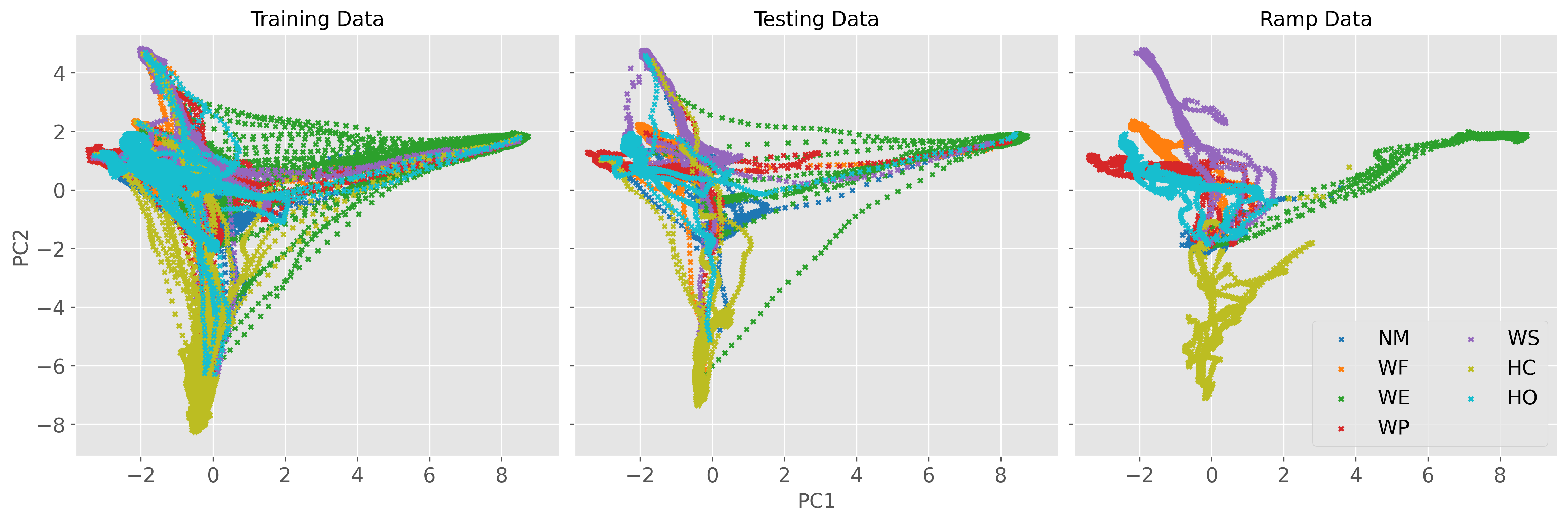}
    \caption{A visualization of the VICReg latent space for an example participant, projected into a 2 dimensional space using PCA. Left: Dynamic Training Data, Middle: Dynamic Testing Data, Right: Ramp Training Data. The colours indicate the labels as determined by our naive labeling scheme.}
    \label{fig:latent_space}
\end{figure*}

\subsection*{Impact of VICReg Pre-training (LSTM-V)}

Figure \ref{fig:latent_space} shows an example latent space obtained from a VICReg trained backbone model. When the embeddings are colour-coded based on their naive labels, it becomes apparent that the latent space formed by the model inherently grouped data points from the same classes together, even without exposure to the labels. Additionally, the model encoded transitions between these groups as narrow corridors. The added dynamics introduced by the continuous transitions can also be clearly seen when comparing between the training data (left) and the ramp data (right).  For example, the training data shows a clear transition can be seen from the (green) WE class to the (cyan) HO class that is absent in the ramp data figure.

To assess the impact of VICReg pre-training, we compared the performance of LSTM-V with LSTM-D and LDA-D. Post-hoc pairwise comparisons revealed significant improvements for LSTM-V compared to the other two on all metrics with a Medium or greater effect size  ($p < 0.0001$; $d > 0.6$ in all cases), except for TOFF. Notably, LSTM-V exhibited better INS and SS-INS with a Huge effect size ($p < 0.0001$, $d > 2.9$). However, it also exhibited a statistically significant worse TOFF, also with a Huge effect size ($p < 0.0001$, $d > 2.0$). This suggests a potential trade-off between improved performance on other metrics and model responsiveness.

\subsubsection*{Overall Improvement Compared to Baseline}

It's important to consider the overall improvement in performance achieved by LSTM-V compared to the baseline model (LDA-R). Here, LSTM-V significantly outperformed LDA-R on all metrics with Very Large or greater effect size ($p < 0.0001$; $d > 1.2$ for all cases) , except for TOFF. This highlights the substantial benefit of using VICReg pre-training and LSTM architecture, particularly when trained on continuous dynamic data.

\subsection*{Impact of Rejection}

\begin{figure*}[htbp]
    \centering    
        \includegraphics[width=0.95\textwidth]{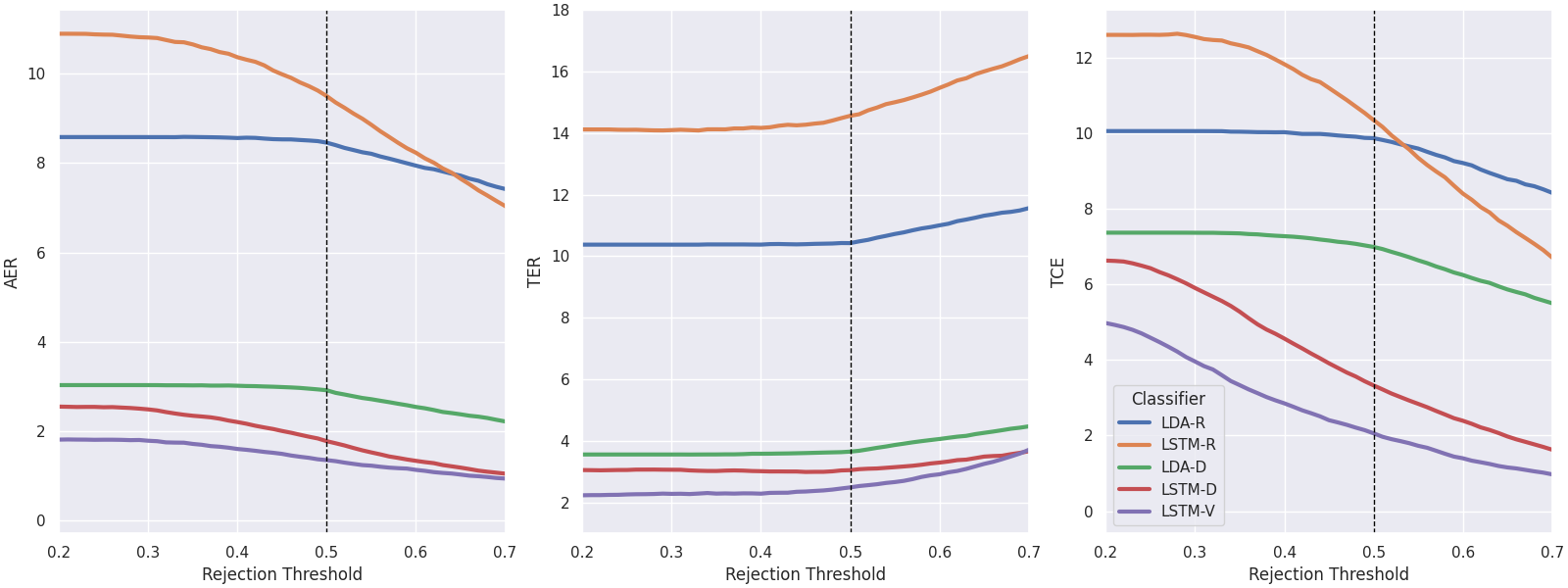}
    \caption{Impact of rejection threshold on average error rates for various models. Thresholds $>0.5$ show an increase in TER, indicating over-rejection. }
    \label{fig:rejection_curves}
\end{figure*}

Figure~\ref{fig:rejection_curves} depicts the relationship between rejection threshold and \emph{average} error rates (AER, TER, and TCE) for the various models. As anticipated, a reduction in AER and TCE was observed across all models as the rejection threshold is increased. However, beyond a certain threshold (approximately $0.5$), a corresponding increase in TER was observed, indicating possible over-rejection. This motivated the choice of setting the rejection threshold to $0.5$.

With the introduction of confidence-based rejection (threshold of $0.5$), the performance difference in AER between LDA-R and LSTM-R becomes non-significant ($p > 0.8$), suggesting rejection can mitigate some of the limitations of LSTM models trained on ramp data. As expected, most metrics improve with rejection, except for TTD. LSTM models consistently benefited more from rejection than LDA models, with LSTM-V demonstrating the greatest advantage. Notably, LSTM-V outperforms LDA-R with a larger effect size ($d > 1.9$) across most metrics, except for TTD where the effect size decreased ($d > 0.6$) but remained significant. Furthermore, with rejection, LSTM-D exhibited a statistically significant decrease in TCE compared to LDA-D with a Huge effect size ($p < 0.001, d > 3$). Interestingly, while LSTM-D and LSTM-V had smaller TTD than LDA-D without rejection, this relationship is reversed with rejection ($p<0.001, d > 0.3$). Nevertheless, it is crucial to note that rejection leads to a more substantial reduction in SS-INS, INS, and TCE for LSTM-D and LSTM-V compared to LDA-D.

\section{Discussions}
Results comparing dynamic data training types with ramp data training types indicated that both LDA and LSTM models exhibit improvements across almost all metrics when trained on continuous dynamic data, even with a naive labeling approach. Including class transitions seems to help both models establish clearer class boundaries. Also, the LSTM outperformed the LDA across almost all metrics when trained with continuous data. These findings echo the results observed in our previous work \cite{raghu_enabling_2024}, where we demonstrated similar performance gains using a GRU-based model. The consistency of these results across different temporal architectures suggests that the benefits of training with continuous dynamic data are not specific to a particular model and may generalize to other temporal classifiers.  While this encouraging, there may be room for improvement. 

First, the current naive labeling strategy may not be sufficient for the LSTM to fully exploit its capabilities in learning dynamic, as it may have provided a poor representation of the structure in the data. The stronger performance of VICReg pre-trained model, which didn't require labeled data during pre-training, supports this notion. There may be a better way to split transitions into preceding and succeeding motion classes, or it may be necessary to include `transition classes' in addition to the motion classes. It may also be possible to label transitions as an `unknown' class rather than as one of the motion classes. Furthermore, the challenge of label noise in general is being actively researched, and techniques such as loss correction \cite{patrini_making_2017}, soft labeling \cite{vyas_learning_2020}, and active learning \cite{ren_survey_2021} offer promising avenues for exploration. Before the limits of a labeling approach can be ascertained when training with continuous dynamic data, more work needs to be done to figure out how to label the data appropriately.

Second, the dataset, while containing transitions, may not have captured the full range of dynamic variability often seen in real-world scenarios. There were likely variations from trial-to-trial in transition speeds, intensities, and durations of steady-states, however a dataset explicitly encompassing a wider array of these dynamics might help the LSTM better learn the nuances in the data.

Results comparing the non-temporal vs temporal model indicated that the LSTM trained with continuous dynamic data performed better than LDA trained with the same data. This was expected because the LSTM can leverage the temporal information in the data. Intriguingly, when the LSTM was trained with ramp data, the performance was generally worse than the LDA. This was unexpected because ramp data still includes some dynamics (from no motion to each motion class). On closer inspection however, we note that it is plausible that the lack of class-to-class transition information during training lead the LSTM to overtune to ramp-like temporal dynamics (where contractions always increase over time), leading to poor performance when classifying continuous dynamic data. Regardless, this outcome suggests that caution is warranted when training deep learning models like LSTMs with conventional ramp training data, which are commonly used in the literature. In addition to these findings, and inline with our previous work \cite{raghu_decision-change_2023}, as shown in Figure \ref{fig:decision_stream}, the LSTM models exhibit a much larger range of posterior probability (i.e., classifier confidence) values compared to the LDA models. This has the potential to make post-processing tools such as rejection \cite{robertson_effects_2019} more effective as rejection relies on good separability of confidence values between correct and incorrect decisions.

Results demonstrate that the VICReg pre-trained model (LSTM-V) outperformed the baseline LDA (LDA-R) and other classifiers trained on continuous dynamic data (LSTM-D and LDA-D) across almost all metrics, both with and without rejection. The only exception was a slightly increased offset time (TOFF) observed in LSTM-V. This lag might be attributed to a combination of factors. The VICReg pre-training, which emphasizes learning noise-invariant representations, could inherently prioritize stability over rapid responsiveness. Additionally, the specific hyperparameter settings and the strength of data augmentations might further contribute to this behavior by encouraging the model to favor smooth transitions over immediate reactions to changes in the input data.

Interestingly, applying rejection led to a decrease in TOFF but an increase in TTD for the LSTM models, particularly LSTM-D and LSTM-V. This observation can be explained by their tendency to exhibit lower confidence values during transitions, as seen in Figure \ref{fig:decision_stream}. With rejection, these low-confidence predictions from the previous steady-state class near the transitions are relabeled to NM class, effectively initiating the `transition' sooner. However, the significant reduction in TCE observed with rejection suggests that this earlier transition onset might not negatively impact overall performance. Nevertheless, further investigation through usability studies is warranted to confirm this hypothesis and assess the practical implications of both VICReg's increased TOFF and the increase in TTD due to rejection.

To gain some insight into why the VICReg pre-trained model yielded better performance, we inspected the embeddings obtained from the backbone. Figure \ref{fig:latent_space} shows that the latent space clusters embeddings by class. This is interesting, because the loss function does not explicitly encourage the model to form class specific clusters, but it is also inline with recent research that has indicated that SSL approaches do cluster semantically similar data together \cite{ben-shaul_reverse_2023}. Even more interesting are the narrow corridors encoding the transitions in the latent space, which hasn't been demonstrated in existing literature. While our motivation for using VICReg was to use it with dynamic continuous data where labeling was ambiguous, it is worth asking if it affords the same performance improvements when used with ramp data. To answer this question, we conducted an investigation where we trained our backbone with ramp data and VICReg loss and found that the model often struggled to converge. To explore why it failed, we projected ramp data onto the latent space learned from training the backbone with continuous data, as shown in Figure \ref{fig:latent_space} (right-most subplot). In the figure, no narrow corridors between classes are apparent, except for those that start from no motion. This underscores a vital point: exposing the model to class-to-class transitio
dynamics is essential to fully develop the learned latent space. This leads us to hypothesize that the absence of these dynamics in the ramp data causes the latent space to collapse, and suggests that integrating more real-world dynamics, like varying speeds and intensities, could result in an even more resilient latent space representation. 

The performance improvements observed with VICReg pre-training stem from the introduction of data augmentations during training. Research indicates that augmentations play a crucial role in achieving optimal performance with SSL methods, often surpassing the impact of choice of loss function   \cite{morningstar_augmentations_2024}. We conducted a preliminary exploration, where we employed one augmentation at a time and found that the time-shift augmentation had more impact than the others. For example, the time-shift augmentation had significantly better INS and TCE compared to the other two augmentations. This suggests that time-related augmentations might be particularly beneficial for sEMG-PR tasks. Nevertheless, the other two augmentations are known to bolster the network's ability to withstand noise in the data, and thus are still beneficial. Regardless, the impact of other augmentation techniques proposed in the wider literature, and the corresponding hyperparameters, remains to be fully explored. There might be alternative augmentation techniques that could significantly enhance downstream classification performance. For example, future augmentations may perhaps focus on anchoring slower transitions to faster ones, rather than simply co-locating them in the latent space, to improve the responsiveness of control. Regardless, further research is necessary to not only scrutinize individual augmentations but also to examine the interaction between various augmentation methods, specifically in the context of sEMG-PR tasks.

In addition to exploring augmentation strategies, future work could also investigate the optimal hyperparameter settings for SSL training. While the current configuration yielded promising results, a more thorough hyperparameter search may uncover further performance gains. Furthermore, although we did not explicitly include a separate projection head in our SSL framework, the final non-linear layers of our network can be considered to serve a similar function. Many SSL approaches discard some of the layers learned during pre-training; however, we retained all layers as this appeared to be effective in our experiments. Future research could systematically investigate the impact of layer discarding on performance. Moreover, while we employed the LSF4 feature set in this study, other feature sets warrant further investigation. Our internal testing suggests that the choice of feature set may not be a critical factor, particularly for deep learning models. Nevertheless, future research could systematically compare different feature sets and explore their interaction with various classifier architectures.

While the offline results are encouraging, online usability studies are crucial to validate performance gains in real-world scenarios. A key anticipated challenge is determining appropriate proportional control mapping for continuous dynamic data. Traditionally, ramp data is used to determine proportional control mapping based on the distribution of amplitudes observed during ramp contractions. These contractions generally provide a rich representation of the distributions.  However, distributions provided by continuous dynamic data may not be as rich, which could make the mapping more challenging. This may simply require a modified threshold selection approach to accommodate the less structured nature of dynamic data collection, but  more needs to be done to fully explore this interaction between proportional control mapping and continuous dynamic data.

Finally, it is crucial to acknowledge that the protocol used in this work inherently resulted in a larger volume of continuous dynamic data compared to ramp data. While this discrepancy could raise questions about whether the observed performance improvements are solely attributable to increased data size, our internal testing suggests otherwise. Preliminary results indicate that similar benefits can be achieved with significantly less continuous dynamic data, even with just one trial for training. Notably, the duration of a single continuous dynamic trial is comparable to the overall ramp data collection time. Moreover, exploring transfer learning or domain adaptation techniques, as suggested in related works \cite{cote-allard_deep_2019, wu_transfer_2023, lehmler_deep_2022}, could further reduce the amount of continuous dynamic data required to achieve comparable performance gains. We are currently conducting further research to fully explore the interplay between data volume, transfer learning, and performance, with the aim of optimizing data collection protocols and ultimately minimizing user burden.

\section{Conclusion}
This works contributes to the body of sEMG-PR by showing: 1) the advantage of training with continuous dynamic data, noting that temporal models better exploit this kind of data, 2) the interaction between model type and training data type, and the caution needed when employing temporal models with ramp data, 3) that SSL with VICReg pre-training further enhances performance with dynamic data, and 4) the effectiveness of confidence-based rejection in improving performance, particularly for LSTM models, with the SSL approach demonstrating the most significant gains. It therefore opens doors for exploring improved labeling strategies, developing deep temporal models that leverage transfer learning for better generalization with limited data, investigating a wider range of time-related and other augmentations for sEMG-PR, and examining their interactions for potential performance gains. It also validates the SSL approach offline, with the need for a forthcoming followup online usability study to test its benefits with the user in the loop.

\section*{Acknowledgments}
This work was supported in part by NSERC Grant 2020-04776, Canada.


\section*{Conflict of interest}
The authors declare no conflicts of interest in relation to this work.

\section*{CRediT authorship contribution statement}
\textbf{Shriram Tallam Puranam Raghu}: Conceptualization, Methodology, Software, Investigation, Formal analysis, Data curation, Writing original draft, Writing - review \& editing, Visualization, Resources. \textbf{Dawn T. MacIsaac}: Conceptualization, Methodology, Software, Investigation, Formal analysis, Data curation, Writing – original draft, Writing - review \& editing, Visualization, Resources. \textbf{Erik J. Scheme}: Conceptualization, Methodology, Software, Investigation, Formal analysis, Data curation, Writing – original draft, Writing - review \& editing, Visualization, Resources, Funding acquisition.

\section*{Declaration of generative AI and AI-assisted technologies in the writing process}
During the preparation of this work, the author(s) used Google Gemini AI to improve clarity, conciseness, and flow. After using this tool, the author(s) reviewed and edited the content as needed and take(s) full responsibility for the content of the 
\printbibliography
\end{document}